\documentstyle[prl,aps,multicol,epsf]{revtex}
\begin{document}
\draft
\widetext
\title{Ground state of a 
double-exchange system containing impurities: bounds of ferromagnetism}
\author{Eugene Kogan$^{1}$, Mark Auslender$^2$ and Eran Dgani$^{1}$}
\address{$^1$ Jack and Pearl Resnick Institute 
of Advanced Technology,
Department of Physics, Bar-Ilan University, Ramat-Gan 52900, 
Israel\\
$^2$ Department of Electrical and Computer Engineering,
Ben-Gurion University of the Negev,
P.O.B. 653, Beer-Sheva, 84105 Israel}
\date{\today}
\maketitle
\begin{abstract}
\leftskip 54.8pt
\rightskip 54.8pt
We study the boundary between
ferromagnetic and non-ferromagnetic ground state  of a 
double-exchange system with  quenched disorder for arbitrary relation
between Hund exchange coupling and electron band width. The boundary 
is found both from the solution of
the Dynamical Mean 
Field Approximation equations and from the comparison of the energies of the
saturated ferromagnetic and paramagnetic states. Both methods give very similar
results. To explain the disappearance of ferromagnetism in part of the parameter
space we  derive from the double-exchange Hamiltonian with
classical localized spins  in the limit of large but finite Hund
exchange coupling  the $t-J$ model (with
classical localized spins).
\end{abstract}
\begin{multicols}{2}
\narrowtext

\section{Introduction}

The problem of ferromagnetism in the double-exchange (DE) 
model \cite{zener,anderson,degennes}, which  appears due to the Hund 
exchange coupling  between the core
spins and the
mobile carriers, has a long history (see Ref. \cite{furukawa2}
and references therein). Frequently the modeling
deals with the case of infinite
Hund exchange. In this approximation the ground state of the system 
is always ferromagnetic (FM). 
To describe possible non-FM ground states of
such a system one has to introduce direct  antiferromagnetic
exchange  between the core spins. However, the finite Hund exchange by
itself generates effective antiferromagnetic exchange,
thus destroying ferromagnetism in a part of the system parameter space. 

In our previous publication \cite{aus02} on the basis of DMFA 
we obtained the closed formula for the
ferromagnet-paramagnet (FM-PM) transition temperature $T_c$
for 
the double-exchange system  for arbitrary relation
between Hund exchange coupling and electron band width.
In this paper we
present a detailed study of  of the boundary between
FM and non-FM ground state  of the 
system found from that formula, and, independently, from comparison of energies
of the saturated FM and PM states. In addition,  from the double-exchange Hamiltonian with
classical localized spins  in the limit of large but finite Hund
exchange coupling we obtain the classical version of the $t-J$ model.

\section{Hamiltonian and DMFA equations}

Consider  the DE model with random on-site energies. The Hamiltonian  of
the model  is 
\begin{eqnarray}
\label{HamDXM}  
H = \sum_{nn'\alpha} t_{n-n'} c_{n\alpha}^{\dagger} c_{n\alpha}+
\sum_{n\alpha}V_n c_{n\alpha}^{\dagger} c_{n\alpha}\nonumber\\ 
-J \sum_{n\alpha\beta} {\bf m}_n\cdot 
{\bf \sigma}_{\alpha\beta}c_{n\alpha}^{\dagger} c_{n\beta},
\end{eqnarray}
where $t_{n-n'}$ is the electron hopping,  $J$ is the effective 
exchange 
coupling between a core spin and a conduction electron,
$\hat{\bf \sigma}$ is the vector of the Pauli matrices, and $\alpha,\beta$ are
spin indices. 
We express the localized (classical) spin  by
 $ {\bf m}_n = (m_n{}^x, m_n{}^y, m_n{}^z)$ 
with the normalization $|{\bf m}|^2 = 1$.
To take into account  the chemical
disorder  introduced by doping impurities, which is generic for 
the manganites and many other DE systems, we consider the case 
random
on-site energy $V_n$.

In a single electron representation the Hamiltonian 
 can be presented as
\begin{equation}
\label{generic}
H_{nn'}=H^0_{n-n'}+\left(V_n-J {\bf m}_n\cdot {\bf \sigma}\right)\delta_{nn'};
\end{equation}
the first  is  translationaly invariant, the second describes
quenched disorder, and the third - annealed disorder.

The DMFA, as applied to the problem under consideration,  is based on two assumptions. The first assumption is 
that the averaged, 
with respect to   random orientation of localized spins and
random on-site energy $V$,  locator 
\begin{eqnarray}
\hat{G}_{\rm loc}(z)= 
\left\langle\hat{G}_{nn}(z)\right\rangle_{{\bf m},V},
\end{eqnarray}
where
\begin{eqnarray}
\label{green}
\hat{G}(z)=(z-H)^{-1}, 
\end{eqnarray}
can be expressed through the  the local self-energy $\hat{\Sigma}$ by the
equation
\begin{eqnarray}
\label{local}
\hat{G}_{\rm loc}(z) =g_0\left(z - \hat{\Sigma}(z)\right),
\end{eqnarray}
where
\begin{eqnarray}
\label{g}
g_0(z) =\frac{1}{N}\sum_{\bf k}\left(z-H^0_{\bf k}\right)^{-1} 
\end{eqnarray}
is the bare (in the
absence of the disorder and  exchange interaction) locator. Thus introduced
self-energy satisfies equation
\begin{eqnarray}
\hat{G}_{\rm loc}(z)=\left\langle \frac{1}
{\hat{G}_{\rm loc}^{-1}(z)+\hat{\Sigma} (z)- 
V_n+J{\bf m}\cdot\hat{\bf \sigma}}\right\rangle_{{\bf m},V}.
\label{cpa}
\end{eqnarray}
The system of equations (\ref{local}) and  (\ref{cpa}) is very much similar to 
the well known CPA equations (see \cite{ziman} and references therein), as
generalized to the case when  the quantities  $\hat{G},\hat{\Sigma}$
and $\hat{g}$ are $2\times 2$ matrices in spin space \cite{kubo}.
The system of equations however, is not yet closed. The averaging with respect
to annealed disorder 
is principally different from the averaging with respect
to quenched disorder.

The second assumption of the DMFA is the prescription for the
determining, in our case, the probability 
of a spin configuration self-consistently with the solutions of
the Eqs. (\ref{local}) and (\ref{cpa}).
To formulate the DMFA equation for this probability, taking into account both
kinds of the disorder, 
let us start from the general formula for
the partition  function 
\begin{equation}
{\cal Z}_{V_n}=\int \exp \left( -{\rm Tr}\sum_s\log \hat{G}(z_s) \right)
\prod_{n}d{\bf m}_{n},
\label{grandz}
\end{equation}
where  $z_s = i\omega_s + \mu$;  $\omega_s$ is  the Matsubara frequency 
and $\mu$ is the chemical potential. 
The averaging over $\left\{{\bf m}_n \right\}$ is given by
\begin{eqnarray}
\left\langle \Phi \right\rangle _{\bf m}=\frac{1}{{\cal Z}_V}\int
\exp \left(-{\rm Tr}\sum_s\log \hat{G}(z_s)\right)
\Phi({\bf m})\prod_{n}d{\bf m}_{n}.
\label{funav}
\end{eqnarray}
All observables, in particular thermodynamic potential $\Omega$, should 
additionally be
averaged over the realizations of the quenched disorder; in particular
\begin{equation}
\Omega=-\frac{1}{\beta }\left\langle \log {\cal Z}_V
\right\rangle _{{\bf m},V}.  
\label{freenergy}
\end{equation}
The DMFA approximates the multi-spin probability 
${\cal Z}_V^{-1}\exp \left(-{\rm Tr}\log
\hat{G}\right)$ as a product of one-site probabilities in such a way, that
\begin{equation}
\frac{\delta \Omega}{\delta \hat{G}_{\rm loc}}=0.
\end{equation} 
The result for the one-site probability reads (for details of the
calculation see Ref. \cite{ak2}):
\begin{eqnarray}
\label{prob2}
P_{V_n}({\bf m})\propto \exp\left[-\beta\Delta\Omega_{{\bf m},V_n}\right],
\end{eqnarray}
where 
\begin{eqnarray}
\label{probability}
\Delta\Omega_{{\bf m},V_n}= -\frac{1}{\beta}\sum_s {\rm Tr} \log 
\left[1+\hat{G}_{\rm loc}(z_s)\right.\nonumber\\
\left.\left(
J{\bf m}\cdot\hat{\bf \sigma}- 
V_n+\hat{\Sigma} (z_s)\right)\right] e^{i\omega_s 0_{+}}.
\end{eqnarray}
is the change of the  thermodynamic potential 
of the electron
gas described by the Green's function $\hat G_{\rm loc}$ 
due to interaction with a single impurity \cite{doniach,chat}.

The right hand side of Eq. (\ref{prob2}),  
is a complicated non-linear functional of $P_V({\bf m})$.
However, 
 if we
are interested only in the transition
temperature $T_{c}$,  the problem can be reduced to a
traditional mean field (MF) equation.
In   linear  with
respect to magnetization $M$ approximation   Eq. (\ref{prob2}) takes the form
\begin{eqnarray}
P_{V_n}({\bf m})\propto  \exp\left( -\beta I_{V_n}{\bf M}\cdot{\bf m}\right).
\label{probability2}
\end{eqnarray}
Non-trivial solution of the MF equation 
\begin{equation}
{\bf M}=\int \left\langle P_{V_n}({\bf m})\right\rangle_V {\bf m}d{\bf m}.
\end{equation}
 can exist only for 
$T<T_{\rm c}$, where $T_c=\frac{1}{3}\left\langle I_{V_n}\right\rangle_V$.

\section{$T_{\rm c}$ for the semi-circular DOS}

For simplicity consider the semi-circular (SC) bare density of states (DOS)
$N_0(\varepsilon)$,  the bandwidth being $2W$.
Then
\begin{eqnarray}
\label{gint}
g_0(z)=\int \frac{N_0(\varepsilon)d \varepsilon }{z - \varepsilon}
=\frac{2}{W}\left[\frac{z}{W}-
\sqrt{\left(\frac{z}{W}\right)^{2}-1}\right].
\end{eqnarray}
For this case
\begin{equation}
\label{sigma}
\hat{\Sigma}=z-2w\hat{G}_{\rm loc}-\hat{G}_{\rm loc}^{-1},
\end{equation}
where $w= W^2/8$.
Thus from Eqs. (\ref{local}) and (\ref{cpa}) we  obtain a single equation
 for $\hat{G}_{\rm loc}$
\begin{equation}
\label{sigma2}
\hat{G}_{\rm loc}(z)=\left\langle\frac{1}{z-2w\hat{G}_{\rm loc}(z)- V_n
+J{\bf m}\cdot\hat{\bf \sigma}}\right\rangle_{{\bf m},V},
\end{equation}
and Eq. (\ref{probability}) can be presented as
\begin{eqnarray}
\label{probability4}
\Delta\Omega_{{\bf m},V_n}= \frac{1}{\beta}\sum_s \log \det
\left[z_s-2w\hat{G}_{\rm loc}(z_s)\right.\nonumber\\
\left.- V_n
+J{\bf m}\cdot\hat{\bf \sigma}\right] e^{i\omega_s 0_{+}}.
\end{eqnarray}
In linear with respect to $M$ approximation
\begin{equation}
\hat{G}_{\rm loc}= g \hat I -
hJ{\bf M\cdot \hat\sigma},
\end{equation}
where $g$ is locator in paramagnetic phase, given by the equation
\begin{eqnarray}
\label{rq2}
g=\frac{1}{2}\left[\left\langle\frac{1}{z-2wg-V_n-J}\right\rangle_{V}\right.
\nonumber\\
\left.+ \left\langle\frac{1}{z-2wg-V_n+J}\right\rangle_{V}\right],
\end{eqnarray}
and the quantity $h$  is given by the formula
\begin{eqnarray}
\label{hv}
h = \frac{\left\langle \Delta_{V_n}\right\rangle_V}
{1 -\frac{4J^2w}{3}\left\langle\Delta_{V_n}^2\right\rangle_V 
-2w\left\langle\Delta_{V_n}\right\rangle_V},
\end{eqnarray}
where  
\begin{equation}
\Delta_{V_n}(z_s)=\frac{1}{\left[z_s-2wg(z_s)-V_n\right]^2 -J^2}.
\end{equation}
Expanding Eq. (\ref{probability4}) we obtain
the effective exchange integral  is   
\begin{eqnarray}
I_{V_n}= \frac{4J^2w}{\beta}\sum_s h(z_s)\Delta_{V_n}(z_s).
\label{probability3}
\end{eqnarray}
If we transform the sum
over the imaginary Matsubara frequencies in the right-hand side of 
Eq. (\ref{probability3}) to integral over real energies $E$, we obtain for the
$T_{\rm c}$  
\begin{eqnarray}
\label{Theta}
T_{\rm c}=\frac{4J^2w}{3\pi}\int_{-\infty}^{\infty}f(E)
\mbox{Im}\left[h(E_+)\left\langle \Delta_{V_n}(E_+) \right\rangle_V\right]dE,	
\end{eqnarray}
where $f(E)$ is the Fermi function, and $E_+= E+i0$.
To the best of our knowledge this closed formula which takes into account both
finite value of Hund exchange and quenched disorder was obtained and studied 
by us for the
first time \cite{aus02}.  
According to this formula the boundary between the FM and non-FM
ground states is  found from equating $T_c$ to zero, which reads
\begin{eqnarray}
\int_{-\infty}^{E_F^{(p)}}
\mbox{Im}\left[h(E_+)\left\langle \Delta_{V_n}(E_+) \right\rangle_V\right]dE=0;	
\end{eqnarray}
the Fermi energy $E_F^{(p)}$ is found from the equation
\begin{equation}
n=-\frac{2}{\pi}\int_{-\infty}^{E_F^{(p)}}{\rm Im}\; g(E_+)\; dE,
\end{equation}
where $n$ is the number of electrons per site.

Thus found 
$T_c$ is the temperature at which the paramagnetic state became
unstable with respect to small spontaneous magnetic moment. 
This becomes especially
obvious, if we obtain the  equation for the 
$T_c$  by constructing Landau functional \cite{ak2}.
Independently, one may consider  another 
approach to the same problem
based on the comparison of the energies of the saturated FM state and PM state.
The energy of the  PM state is found from the equation 
\begin{equation}
E_p=-\frac{2}{\pi}\int_{-\infty}^{E_F^{(p)}}E\;{\rm Im}\; g(E_+)\; dE;
\end{equation}
the energy of the saturated FM state and the appropriate Fermi energy are found from 
the equations: 
\begin{eqnarray}
E_f=-\frac{1}{\pi}\int_{-\infty}^{E_F^{(f)}}E\;[{\rm Im}\;
 g_{\uparrow}(E_+)+ {\rm Im}\;
g_{\downarrow}(E_+)]\;dE\nonumber\\
n=-\frac{1}{\pi}\int_{-\infty}^{E_F^{(f)}}{\rm Im}\;[{\rm Im}\;
 g_{\uparrow}(E_+)+ {\rm Im}\;
g_{\downarrow}(E_+)]  dE,
\end{eqnarray}
where the FM locator $g_{\uparrow\downarrow}$ is given by the equation.
\begin{eqnarray}
g_{\uparrow\downarrow}=
\left\langle\frac{1}{z-2wg-V_n\mp J}\right\rangle_{V}.
\end{eqnarray}
The boundary of the
ferromagnetic region at the phase diagram can be
found from the equality between the
two energies: $E_p=E_f$. We thus use an alternative to the second assumption of
the DMFA.

\section{FM -- non-FM boundary}

The boundary of the FM ground state
of the system in case of no quenched disorder is presented on Fig. 1.
\begin{figure}
\epsfxsize=2.5truein
\centerline{\epsffile{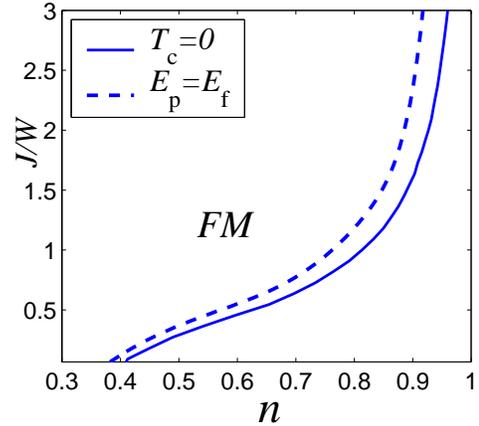}}
\caption{The FM -- non-FM boundary for the case of no quenched disorder
in the coordinates of
relative strength of the Hund exchange $J/W$ and electron concentration $n$.} 
\end{figure}
The diagram agrees with those obtained on the basis 
of numerical calculations \cite{dag} and
from qualitative reasoning \cite{chat}.

To consider the influence of chemical disorder
we consider the model  in which $V_n=V$ with the probability $x$,
and
$V_n=0$ with the probability $1-x$, thus $x$ being the concentration of
impurities. In this model we consider the number of electrons $n$ and the
concentration of impurities $x$ as two independent parameters.
 Solving equation for the locator  we obtain the boundaries
 which are
presented on Figs. 2-4. It is interesting that ferromagnetism is now precluded
in much larger region of
the $J/W-n$ plane. 
\begin{figure}
\epsfxsize=2.5truein
\centerline{\epsffile{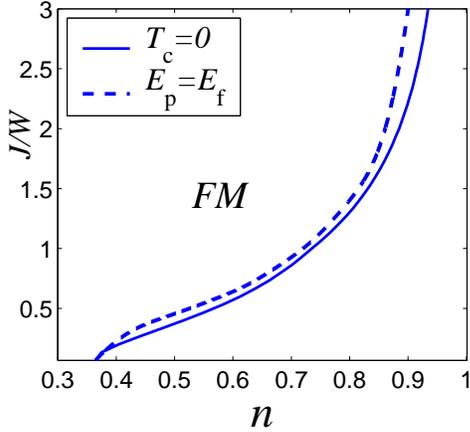}}
\caption{The FM -- non-FM boundary for $V/W=.5$ and $x=.3$
in the coordinates of
relative strength of the Hund exchange $J/W$ and electron concentration $n$.} 
\end{figure}
\begin{figure}
\epsfxsize=2.5truein
\centerline{\epsffile{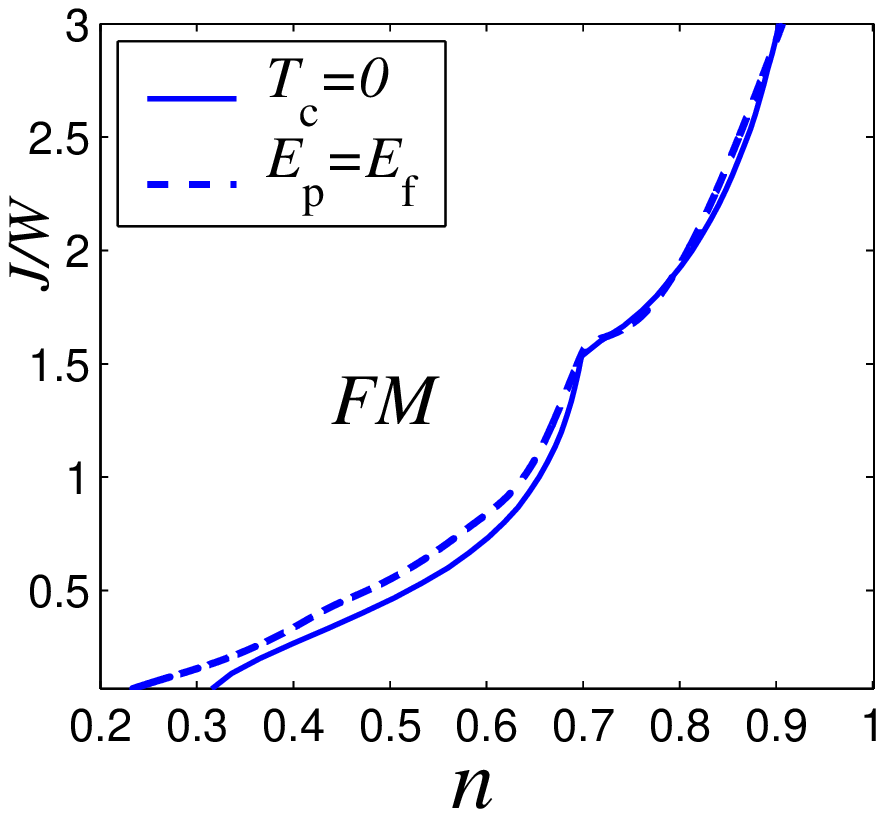}}
\caption{The FM -- non-FM boundary for $V/W=1$ and $x=.3$
in the coordinates of
relative strength of the Hund exchange $J/W$ and electron concentration $n$.} 
\end{figure}
\begin{figure}
\epsfxsize=2.5truein
\centerline{\epsffile{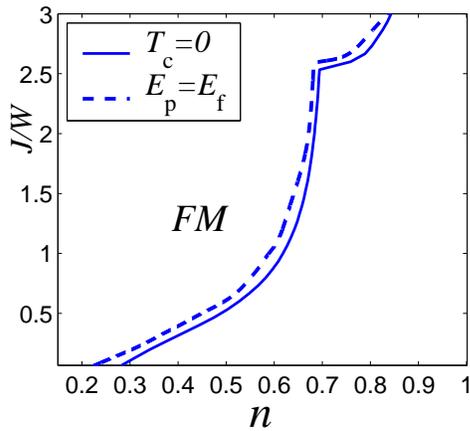}}
\caption{The FM -- non-FM boundary for $V/W=2$ and $x=.3$
in the coordinates of
relative strength of the Hund exchange $J/W$ and electron concentration $n$.} 
\end{figure}
The boundaries between the phases make sharp bends   for 
static disorder strong enough to create a gap between  conduction and
impurity bands both in ferromagnetic and paramagnetic phases \cite{auskog}.
In this case the number of states in the impurity band is equal to the
concentration of impurities, and
for $n=1-x$  the Fermi energy is in
the gap. To emphasize this fact 
we plot the density of states both for FM and PM
phases  at the values of parameters $V/W$, $x$ and $J/W$  which correspond to
the sharp bends on Figs. 3 and 4.
\begin{figure}
\epsfxsize=1.8truein
\centerline{\epsffile{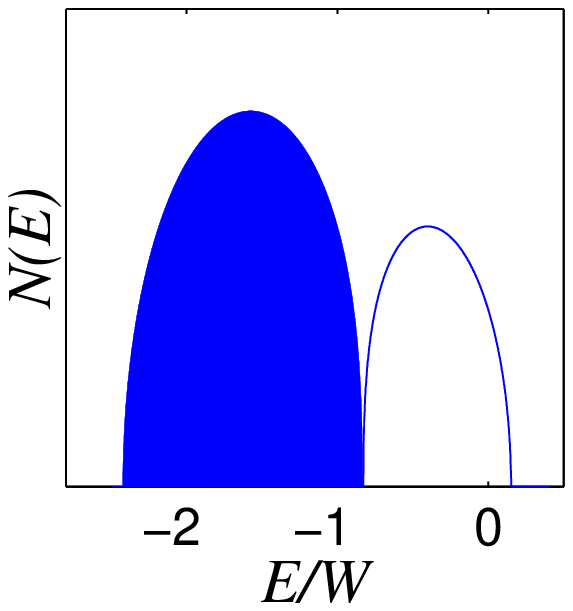}\epsfxsize=1.8truein\epsffile{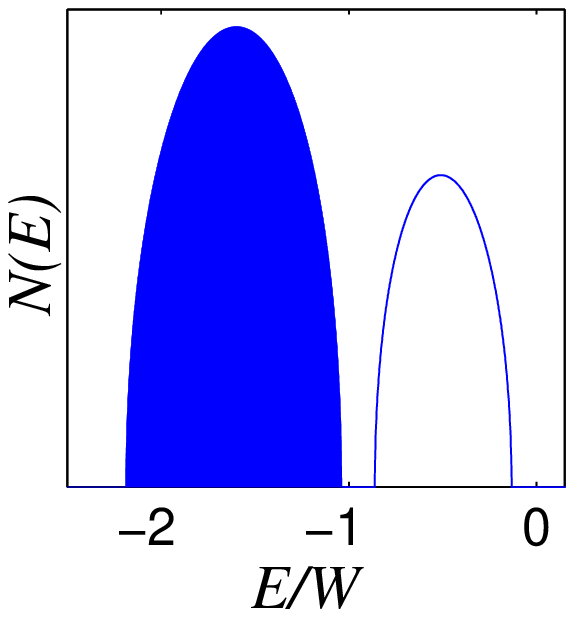}}
\caption{Density of states for $V/W=1$, $x=.3$, $J/W=1.53$ for 
saturated FM phase (left) and PM phase (right); 
occupied states for $n=0.7$ are
 filed.} 
\end{figure}
\begin{figure}
\epsfxsize=1.8truein
\centerline{\epsffile{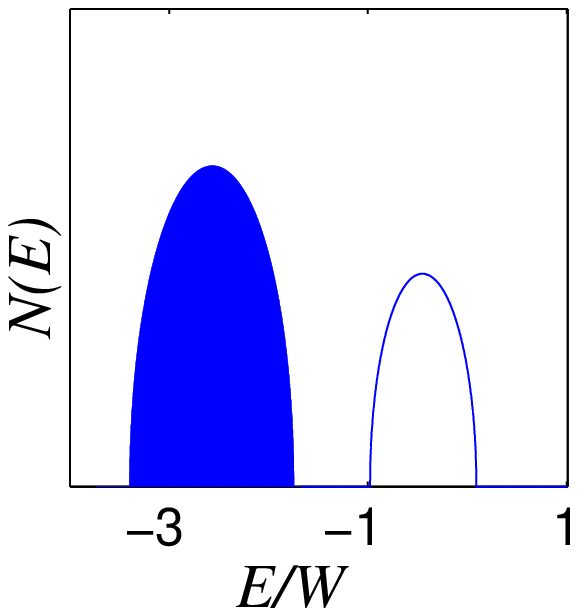}\epsfxsize=1.8truein\epsffile{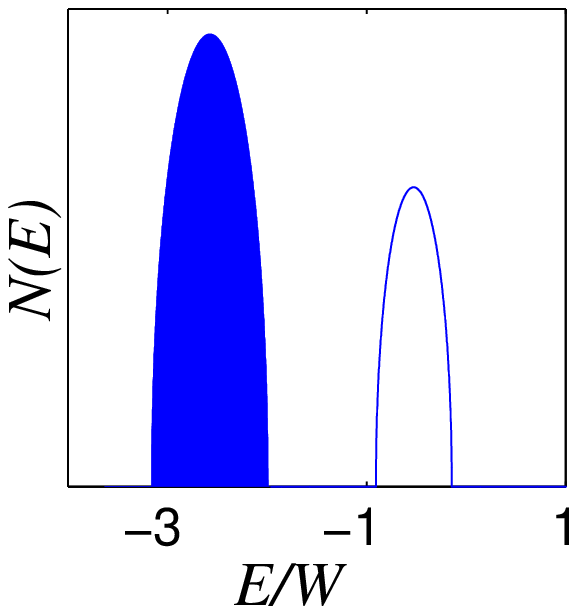}}
\caption{Density of states for $V/W=2$, $x=.3$, $J/W=2.53$ for 
saturated FM phase (left) and PM phase (right); 
occupied states for $n=0.7$ are
filed.} 
\end{figure}

\appendix
\section{From the double-exchange Hamiltonian to the  $t-J$ model: Classical
spins}

The Hamiltonian  of the DE model \cite{zener,anderson,degennes} is 
\begin{eqnarray}
\label{hamiltonian}   
H=-J_H \sum_{n\alpha\beta} {\bf m}_n\cdot 
{\bf \sigma}_{\alpha\beta}c_{n\alpha}^{\dagger} c_{n\beta}-\sum_{nn'\alpha}
t_{nn'} c_{n\alpha}^{\dagger} c_{n'\alpha},
\end{eqnarray}
where $t_{n-n'}$ is the electron hopping,  $J_H$ is the effective Hund
exchange 
coupling between a core spin and a conduction electron,
$\hat{\bf \sigma}$ is the vector of the Pauli matrices, and $\alpha,\beta$ are
spin indices. We express the localized (classical) spin  by
a unit vector whose orientation is determined by polar angle $\theta$ and
azimuthal angle $\phi$. In a single electron representation the Hamiltonian 
 can be presented as
\begin{eqnarray}
\label{hamiltonian2}  
H_{nn'} = H^{ex}+H^{kin}= -J_H {\bf m}_n\cdot {\bf \sigma}\delta_{nn'}- t_{nn'}.
\end{eqnarray}
We consider the case of strong exchange: $J_H\gg W$ where $W$ is the electron
band
width. In this case we should first diagonalize  the exchange part of the
Hamiltonian. This is done by choosing
local spin quantization axis on each site in the direction of ${\bf m}$. 
In this
representation the Hamiltonian  is \cite{kogan}
\begin{eqnarray}
\label{hamlocquant}  
H_{nn'} =  -J_H\sigma^z\delta_{nn'}
-t_{nn'}\left(\begin{array}{cc}a_{nn'} & b_{nn'}\\
b_{n'n}^* & a_{n'n}\end{array}\right),
\end{eqnarray}
where
\begin{eqnarray}
a_{nn'} =  \cos\frac{\theta_n}{2}\cos\frac{\theta_{n'}}{2}
+\sin\frac{\theta_n}{2}\sin\frac{\theta_{n'}}{2}e^{i(\phi_{n'}-\phi_{n})}\nonumber\\
b_{nn'}=\sin\frac{\theta_n}{2}\cos\frac{\theta_{n'}}{2}e^{-i\phi_{n}}-
\cos\frac{\theta_{n}}{2} \sin\frac{\theta_{n'}}{2}e^{-i\phi_{n'}}.
\end{eqnarray}
The transformation to local spin  quantization axis including an additional
Euler rotation angle, which leads to a more involved effective Hamiltonian than 
\ref{hamlocquant},  was introduced by Nagaev \cite{nagaev}.

The next step, like it is
done in the derivation of the $t-J$ model from the Hubbard model \cite{izyumov},
is to apply a canonical transformation 
\begin{eqnarray}
H\rightarrow \tilde{H}=e^SHe^{-S}=H+[S,H]+\frac{1}{2}[S[S,H]]+\dots
\end{eqnarray}
which excludes 
all band-to-band transitions. This can be achieved if we chose the operator $S$
in the form
\begin{eqnarray}
S_{nn'}=\frac{t_{nn'}}{2J_H}\left(\begin{array}{cc}a_{nn'} & b_{nn'}\\
-b_{n'n}^* & a_{n'n}\end{array}\right).
\end{eqnarray}
We have
\begin{eqnarray}
[S,H^{ex}]_{nn'}=t_{nn'}\left(\begin{array}{cc}0 & b_{nn'}\\
b_{n'n}^* & 0\end{array}\right)\nonumber\\
\left[S,[S,H^{ex}]\right]_{nn}
=2\sum_{n''}J_{nn''}\left(\begin{array}{cc}
 -|b_{nn''}|^2 & 0\\
0 & |b_{nn''}|^2\end{array}\right),
\end{eqnarray}
where $J_{nn''}=|t_{nn''}|^2/(2J_H)$.
Keeping terms up to the second order with respect to $t$ (and only site-diagonal
part of the second order terms) we obtain 
\begin{eqnarray}
\label{hamlo}  
\tilde{H}_{nn'} =  -J_H\sigma^z\delta_{nn'}
-t_{nn'}\left(\begin{array}{cc}a_{nn'} &0\\
0 & a_{n'n}\end{array}\right)\nonumber\\
+\sum_{n''}J_{nn''}({\bf m}_n\cdot{\bf m}_{n''}-1)\sigma^z\delta_{nn'},
\end{eqnarray}

In the second quantization form the Hamiltonian (\ref{hamlo})  has the form
(ignoring the constant term)
\begin{eqnarray}
\label{ham}
\tilde{H}=-\sum_{nn'}t_{nn'}a_{n'n}d_n^{\dagger}d_{n'}\nonumber\\
+\sum_{nn'}J_{nn'}{\bf
m}_n\cdot{\bf m}_{n'}(1-d_n^{\dagger}d_{n}),
\end{eqnarray}
where for the case of
hole doping $d^{\dagger}(d)$ is the operator of creation (annihilation) 
of the hole, and for the case of electron doping
it is the operator of creation (annihilation) of the electron.

Looking at the  Hamiltonian obtained, we see that large but finite
Hund exchange dynamically generates antiferromagnetic exchange 
(the second term in Eq. (\ref{ham}). This term, however, 
is not independent upon the electron (hole) subsystem. In this regard this
Heisenberg like term resembles the first non-Heisenbergian term (kinetic
exchange).

This research was supported by the Israeli Science Foundation administered
by the Israel Academy of Sciences and Humanities and BSF.

\end{multicols}
\end{document}